\documentclass{desyproc}
\begin{document}
%------------------------------------
\title{Theoretical prospects for directional WIMP detection}

%for single authors the superscripts are optional
\author{{\slshape  Ciaran A. J. O'Hare}%, Julien Billard$^{2}$, Enectali Figueroa-Feliciano$^{3}$, Anne M. Green$^{1}$, Louis E. Strigari$^{4}$}
\\[1ex]
School of Physics and Astronomy, University of Nottingham, University Park, Nottingham, NG7 2RD, UK}%\\
%$^2$IPNL, Universit\'e de Lyon, France\\
%$^3$Massachusetts Institute of Technology, Cambridge, MA, USA\\
%$^4$Texas A \& M University, College Station, TX, USA}

% if the proceedings are available online (e.g. at Indico)
% please enter the contribution ID or file_name below for the DOI
%\contribID{32}
\contribID{ohare\_ciaran}

% TO THE CONFERENCE EDITORS: 
% please update the following information      
% before sending the template to the authors
\confID{11832}  % if the conference is on Indico uncomment this line
\desyproc{DESY-PROC-2015-02}
\acronym{Patras 2015} % if you want the Acronym in the page footer uncomment this line
%\doi  % if there is an online version we will register DOIs

\maketitle

\begin{abstract}
 	Direct detection of dark matter with directional sensitivity is a promising concept for improving the search for weakly interacting massive particles. With information on the direction of WIMP induced nuclear recoils one has access to the full 3-dimensional velocity distribution of the local dark matter halo and thus a potential avenue for studying WIMP astrophysics. Furthermore the unique angular signature of the WIMP recoil distribution provides a crucial discriminant from neutrinos which currently represent the ultimate background to direct detection experiments.
\end{abstract}

\section{Introduction}
The search for WIMPs by direct detection has in principle a strong directional signature. The motion of the Solar system within the non-rotating dark matter halo of the Milky Way gives rise to an apparent wind of WIMPs coming from a particular direction in the sky aligned with the constellation of Cygnus. The detection of the direction of laboratory-based nuclear recoils consistent with this predicted direction would hence be a smoking gun for the scattering of a particle with Galactic origin. If achievable at the scale of current non-directional experiments, directional detection would not only provide another way of making competitive exclusion limits on the WIMP parameter space but also for the discovery of an unequivocal WIMP signal~\cite{Billard:2011zj}. Beyond this, directional detection is also a novel technique for studying the astrophysics of WIMPs as it probes the full local velocity distribution; without directional information one only has access to the 1-dimensional speed distribution. 

Another area in which directional detection is promising is in the subtraction of the encroaching ``irreducible background'' due to coherent neutrino-nucleus scattering (CNS). For example a 1 keV threshold Xenon detector with a mass of 1 ton operated for a year will detect around 100 neutrino events from $^8$B decay in the Solar core. Given that neutrinos cannot be shielded against they represent the ultimate background to WIMP direct detection experiments. The limiting cross-section at which the neutrino background becomes important is known as the neutrino floor.

The content of this section of the proceedings has been drawn from Refs.~\cite{O'Hare:2015mda, O'Hare:2014oxa} in which further results and technical details can be found.

\section{Directional detection}
The rate of WIMP-nucleus elastic scattering events in the laboratory frame is a function of recoil energy, recoil direction and time. We can write the triple differential recoil rate per unit detector mass for spin-independent interactions with a single nucleus type of mass number $A$ as~\cite{Gondolo:2002np},
\begin{equation}
\frac{\mathrm{d}^3R}{\mathrm{d}E_r\mathrm{d}\Omega_r\mathrm{d}t} = \frac{\rho_0\sigma_{\chi-n}}{4\pi m_{\chi}\mu^2_{\chi n}\Delta t} A^2 F^2(E_r)\int \delta(\textbf{v}\cdot\hat{\textbf{q}}-v_\textrm{min})\,f(\textbf{v}+\textbf{v}_\textrm{lab}(t))\textrm{d}^3 v \, ,
\label{eq:WIMPdirectionalrate}
\end{equation}
where $\rho_0 = 0.3$ GeV cm$^{-3}$ is the local astrophysical density of WIMPs, $\sigma_{\chi-n}$ is the spin-independent WIMP-nucleon cross-section, $m_\chi$ is the WIMP mass, $\mu_{\chi n}$ is the WIMP-nucleus reduced mass and $\Delta t$ is the exposure time of the experiment. The function $F(E_r)$ is the nuclear form factor which describes the loss of coherence in the WIMP-nucleus interaction at high momentum transfer. The velocity distribution, $f(\textbf{v})$, enters as its Radon transform and has been boosted into the laboratory frame by the time dependent lab velocity $\textbf{v}_\textrm{lab}(t)$. The angular dependence of the event rate is a dipole anisotropy peaking towards -$\textbf{v}_\textrm{lab}$.

The unique advantage given by directional information is the potential to make a WIMP ``discovery'' i.e., to claim that a detected particle is of Galactic origin. Once the initial assumption of isotropic backgrounds has been rejected which requires around $\mathcal{O}(10)$ events~\cite{Green:2006cb}, a discovery can be made by checking the consistency of the direction of nuclear recoils with the direction of Solar motion. This can be done with either non-parametric tests on spherical data or with a likelihood analysis and requires as few as $\mathcal{O}(30)$ events~\cite{Billard:2011zj,Green:2010zm}.

Once dark matter has been discovered the search enters the post-discovery phase when it becomes possible to study phenomena regarding the WIMP interaction and perform essentially ``WIMP astronomy'' by observing the velocity distribution of the local dark matter halo. An example of such a study is the detection of tidal streams of dark matter~\cite{O'Hare:2014oxa}. The hierarchical formation of the Milky Way is expected to give rise to a number of streams of dark matter matter wrapping around the Galaxy due to the tidal stripping of material from smaller satellite galaxies~\cite{Purcell:2012sh}. We have found that for a reasonably forecasted directional detector with a CF$_4$ target, 30 kg-yr exposure and threshold of 5 keV, a stream such as that expected from the Sagittarius dwarf galaxy would be detectable with either Bayesian parameter inference or a modified profile likelihood ratio test between a substructure free halo model and one containing a stream~\cite{O'Hare:2014oxa}.

\subsection{Experiments}
Directional detection is a very exciting prospect theoretically but in practice is fraught with experimental limitations. The standard approach with low pressure gas time projection chambers (TPCs) suffers from complications such as limited sense recognition and large angular resolution, as well being inherently low in mass. In light of these restrictions it has become pertinent to consider possible solutions; for instance by compromising on the full 3-dimensional recoil track reconstruction. A 2-d readout for example would be obtained in a gas-TPC without time sampling the anode. A 1-d readout on the other hand consists of only measuring the projection of the recoil track onto the drift direction and currently has no experimental implementation. A 1-d readout strategy has put forward in concept however using dual-phase liquid noble detectors exploiting a potentially measurable directional effect due to columnar recombination~\cite{Nygren:2013nda}. The advantage of such a technique is that it is possible to perform with existing technology and in the liquid phase making it much more readily scalable to higher detector masses.

\section{Directional detection and the neutrino floor}
To follow Eq.~(\ref{eq:WIMPdirectionalrate}) we can write the triple differential recoil rate per unit detector mass for coherent neutrino-nucleus scattering as the convolution of the double differential cross-section and the neutrino directional flux,
\begin{equation}
\frac{\textrm{d}^3 R}{\textrm{d}E_r \textrm{d}\Omega_r \textrm{d}t} =  \frac{1}{m_N} \int_{E_\nu^{\rm min}} \frac{\textrm{d}^2 \sigma}{\textrm{d}E_r \textrm{d}\Omega_r}\times\frac{\textrm{d}^3 \Phi}{\textrm{d}E_\nu \textrm{d}\Omega_\nu \textrm{d}t} \textrm{d}E_\nu \textrm{d}\Omega_\nu \,,
\end{equation}
where $E_\nu^{\textrm{min}}$ is the minimum neutrino energy required to generate a recoil of energy $E_r$ and $m_N$ the nucleus mass. The neutrino directional flux is dependent on the type of neutrino under consideration. For Solar neutrinos the flux is a delta function in direction with a cosine modulation in time due to the eccentricity of the Earth's orbit. For DSNB and atmospheric neutrinos the flux can be approximated as isotropic and constant in time.

\begin{figure}[t]
\begin{center}
\includegraphics[width=0.49\textwidth]{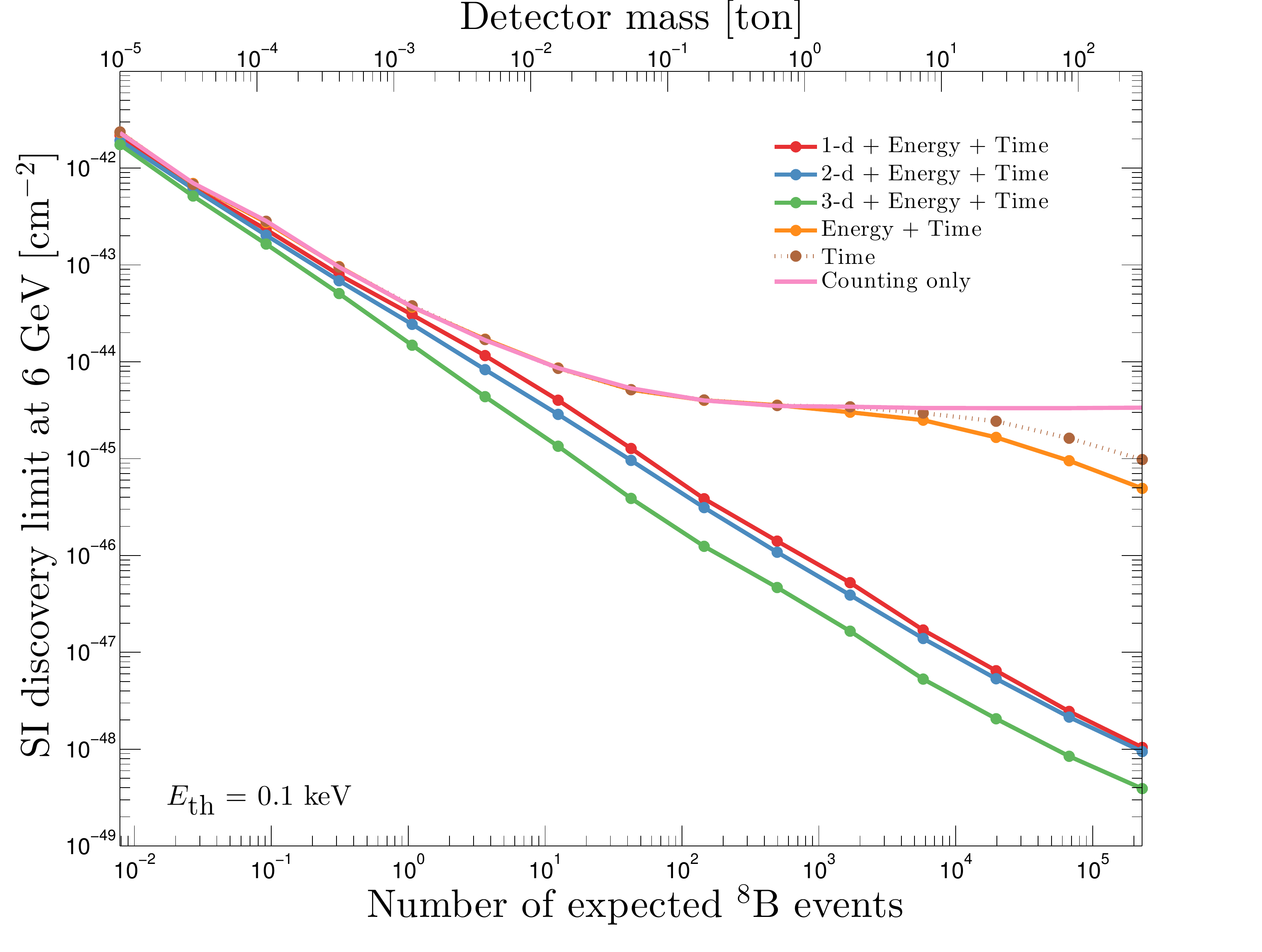}
\includegraphics[width=0.49\textwidth]{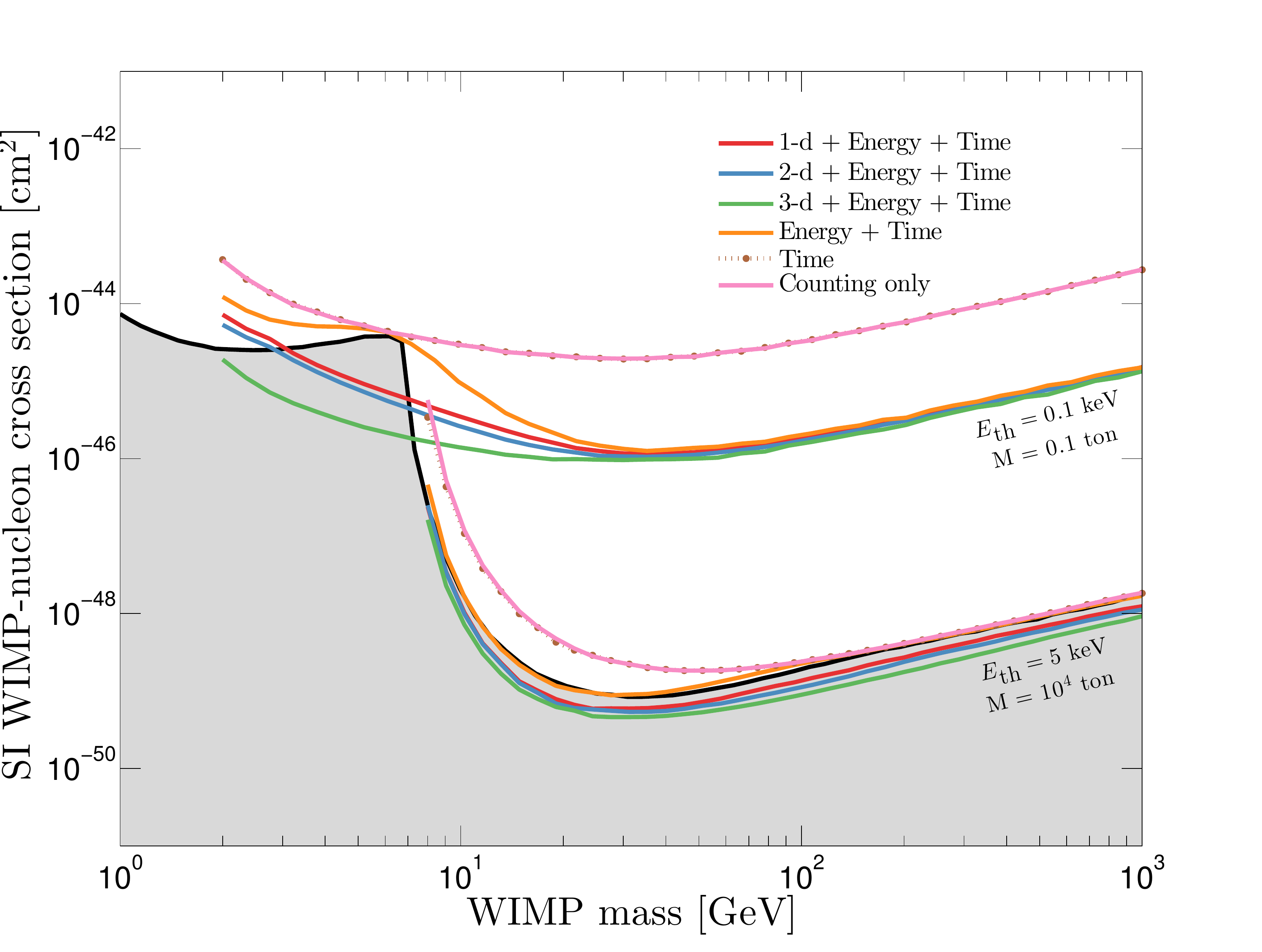}
\caption{{\bf Left:} The dependence of the discovery limit for the spin independent WIMP-nucleon cross-section, $\sigma_{\chi-n}$,  on the mass of a ${\rm Xe}$ detector operated for $1 \, {\rm year}$ using  (from top to bottom)  number of events only (pink line), time information (brown dotted),  energy \& time (orange), energy \& time  plus 1-d (red), 2-d (blue) and 3-d (green) directionality. {\bf Right:} Discovery limits as a function of WIMP mass for fixed detector mass. The upper set of curves correspond to the limits obtained by a detector with a threshold of 0.1 keV and a mass of 0.1 ton, and the bottom set of curves for a 5 keV threshold detector with a $10^4$ ton mass. The shaded region indicates the neutrino floor from Ref.~\cite{Billard:2013qya}.} 
\label{fig:ohare_ciaran_fig1}
\end{center}
\end{figure}

Figure \ref{fig:ohare_ciaran_fig1} shows the discovery limits for the spin-independent WIMP-nucleon cross-section obtained in a Xenon detector located in the Modane underground lab, operated for 1 year. The discovery limit is defined as the minimum cross-section for which 90\% of hypothetical experiments can make a 3$\sigma$ discovery~\cite{Billard:2011zj} and are calculated using the standard profile likelihood ratio test. In the left hand plot we show the limits for a 6 GeV WIMP (which has a recoil spectrum closest to that of $^8$B neutrinos) as a function of detector mass for the 6 readout strategies. Firstly we have a counting only experiment, in such an experiment only the number of events above some energy threshold is measured. In this case the discovery limit plateaus at a value controlled by the $^8$B neutrino flux uncertainty (around 15\%). Including time information to a counting search only improves the discovery limit at very large detector masses/exposures due to the small amplitudes of the annual modulation effects. An energy + time experiment has a slight advantage due to the small differences in the tails of the WIMP and neutrino recoil spectra. Again, as this is a very small effect the detector masses needed to go beyond the neutrino floor are extremely large by directional detection standards. Including 1-d, 2-d and 3-d information the discovery limit cuts below the neutrino floor and retains a $1/M$ scaling due to the significant differences between the angular signatures of the WIMP and neutrino induced recoils. This proves that indeed directional information is a powerful tool for subtracting the Solar neutrino background.

The right hand plot in Fig.~\ref{fig:ohare_ciaran_fig1} shows the limits as a function of WIMP mass for two detector set-ups a low threshold-low mass detector (0.1 keV and 0.1 ton) and high threshold-high mass detector (5 keV and 10$^4$ ton). These numbers are well beyond the current and possibly even forseeable future of directional detection however it is important to choose model experiments with a sizable neutrino background so that the advantage of a directional readout can be observed. In both the low and high mass WIMP range we see the directional limits cut below the non-directional neutrino floor, in the low mass range by a few orders of magnitude and in the high mass range by a factor of a few. This is due to the fact that at low masses distinguishing WIMP from Solar neutrino recoils is much easier as they both possess unique directional signatures with little overlap between the two, whereas for distinguishing WIMP from atmospheric or diffuse supernova background neutrino recoils, the isotropic distibution of the latter two means there is much overlap between the two recoil signals and discriminating between the two to the same significance requires more WIMP events.

\section{Summary}
The detection of dark matter with directional information currently presents the most powerful approach for disentangling the WIMP signal from the ultimate neutrino background. The difference between the angular dependence of the neutrino and WIMP recoil spectra make the two signals distinct in a way that their recoil energies alone do not. We have shown the neutrino floor can be circumvented over the full range of WIMP masses, tackling both neutrinos from the Sun as well as atmospheric and diffuse supernova background neutrinos. Furthermore directional detection offers an exciting prospect for WIMP astronomy by observing features of the local velocity distribution such as tidal streams.

% ****************************************************************************
% BIBLIOGRAPHY AREA
% ****************************************************************************

\begin{footnotesize}

\end{footnotesize}

% ****************************************************************************
% END OF BIBLIOGRAPHY AREA
% ****************************************************************************

\end{document}